\documentclass[12pt,draft]{note_style}
\usepackage{epsf,amsmath}
\begin{document}
\title{How to teach Quantum Mechanics}
\author{Oliver Passon \\
Fachbereich Physik, University of Wuppertal\\
Postfach 100 127, 42097 Wuppertal,  Germany\\
E-mail: Oliver.Passon@cern.ch }

\maketitle

\begin{abstract}
\noindent 
In the spirit and style of John S. Bell's well known 
paper on {\em How to Teach Special  Relativity} \cite{bell_how} 
it is argued, that a ``Bohmian pedagogy''provides a very useful tool to 
illustrate the relation between classical and quantum physics and
illuminates  the peculiar features of the latter.
\end{abstract}
The paper by Bell  on {\em How to Teach Special Relativity} introduces 
the subject with the following remark:
\begin{quote}
I have for long thought that if I had the opportunity to teach this subject, I
would emphasize the continuity with earlier ideas. Usually it is the
discontinuity which is stressed, the radical break with more primitive notions
of space and time. Often the result is to destroy completely the confidence of
the student in perfectly sound and useful concepts already
acquired.\footnote{Notes are to be ignored in a first reading}
\end{quote}
In the following Bell gives an account of the Lorentzian interpretation of
relativistic effects, in which the Lorentz transformations are explained by a
dynamical mechanism (including the assumption of an -- though undetectable -- 
aether) rather than derived from postulates as in the 
approach of Einstein\footnote{How closely Bell actually stuck to Lorentz's
  thinking is debatable \cite{brown_pooley} but does not matter too much in our
  context.}. Bell concludes that teaching relativity can benefit from 
what he calls a ``Lorentzian pedagogy'', i.e. a presentation of the Lorentzian
viewpoint, since ``the longer road sometimes gives more familiarity with the
country''.            

Our note tries to translate this conclusion into the context of quantum
theory, where Bohmian mechanics\footnote{Bohmian mechanics is frequently
  referred to as de$\,$Broglie-Bohm theory, since Louis de$\,$Broglie had
  similar ideas already in 1927. David Bohm's work in 1952 \cite{bohm} was done
  independently.} can serve a similar purpose as the Lorentzian aether theory
with respect to special relativity. The similarities between both theories are
in fact striking: (i) like the Lorentz interpretation,  
Bohmian mechanics derives the key ingredient (namely Born's probability rule)
rather than postulating it, (ii) both, the Lorentz interpretation and Bohmian
mechanics, make the same experimental predictions as their respective standard
theories, while postulating undetectable entities, (iii) both were
appreciated by John S. Bell, and finally, (iv) like the Lorentz interpretation,
Bohmian mechanics 
emphasizes the continuity with earlier ideas more than the radical break with
more primitive notions of (in this case) particles and matter. 

\begin{figure}[t]
\begin{center}
\centerline{\epsfxsize=3.0in\epsfbox{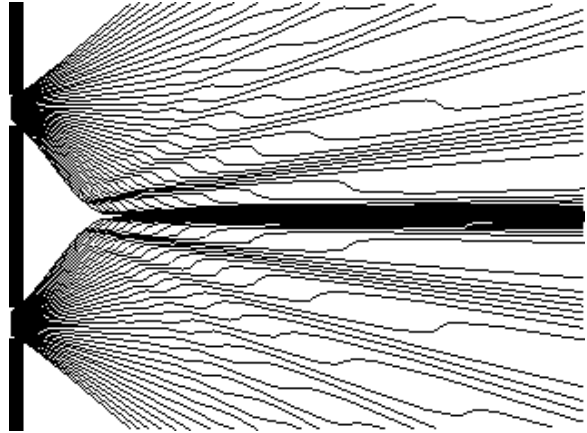}}
\caption[*]{\label{ds} {A numerical simulation of a sample of Bohmian
    trajectories for the double slit experiment \cite{philippidis}. Continuous
    and deterministic trajectories lead to the well 
    known pattern, since they are ``guided'' by the wavefunction,
    which interferes at the double slit.}}
\end{center}
\vspace{-0.8cm}
\end{figure}

Within Bohmian mechanics 
particles keep on moving on trajectories, while the quantum mechanical
interference phenomena come about due to the role of the wavefunction as a
``guiding field''. At the same time this theory is extremely simple! As a
starting point one may take the classical relation between an arbitrary
current ($j$), the related  density ($\rho$) and the velocity field
($v=dx/dt$)\footnote{Bohm's approach \cite{bohm} stressed
  the importance of what he called the {\em quantum potential} -- our
  presentation sticks more to the road favored by Bell
  \cite{bell_on_bohm0,bell_on_bohm,bell_on_bohm2}}: 
\begin{eqnarray}
\label{strom_def}
j = v\cdot \rho 
\end{eqnarray}
In quantum mechanics a so-called probability density and probability current 
is given by the expressions\footnote{Of course our motivation of Bohmian
  mechanics assumes some familiarity with ordinary quantum mechanics and does
  not serve as a first contact to the subject of quantum phenomena.}:
\begin{eqnarray}
\nonumber
\rho&=&|\psi|^2 \\
\label{strom_qm}
j &=&\frac{\hbar}{2mi}\left[ \psi^* (\nabla \psi) - (\nabla \psi^*)
  \psi \right] 
\end{eqnarray}
So it is straight forward to interpret Eqn.~\ref{strom_def} tentatively as an
equation of motion for ``quantum particles'':
\begin{eqnarray}
\nonumber \frac{dx}{dt}&=& \frac{j}{\rho} \\
             &=& \frac{\nabla S}{m} \label{ggbell}
\end{eqnarray}
The last line follows, if one writes $\psi=Re^{\frac{i}{\hbar}S}$ and
substitutes this into the definition of $j$ -- i.e. the phase $S$ is guiding 
the particle motion of a system that is described by the wavefunction $\psi$.
The definition of $j$  is chosen to ensure the following continuity
equation\footnote{In the standard treatment referred to as {\em conservation of
    probability}}:  
\begin{eqnarray}
\frac{\partial \rho}{\partial t}+\nabla j = 0.
\end{eqnarray}
Hence, once the particle density of a system described by the 
wavefunction $\psi$ is $\rho=|\psi|^2$ distributed in position space, it will
stay so. 
In other words: Given a $|\psi(x,0)|^2$ distribution as the initial condition
for 
the particle trajectories, they will produce the very same 
predictions as ordinary quantum mechanics with the probability rule
postulated by Born \cite{born}. One can programme a computer to
integrate Eqn.~\ref{ggbell} for arbitrary solutions of the Schr\"odinger
equation\footnote{Our guiding equation \ref{ggbell} is formulated for the
  1-particle case. The generalization to N-particles and particles with spin
  etc. exists and is straight forward \cite{cushing} }. Let us look at the
famous double slit experiment, which -- as we 
were all told -- allows no explanation in terms of particles moving on
trajectories. The result for some trajectories is displayed in Fig.~\ref{ds}
and shows exactly the quantum mechanical interference pattern since (i) the
wavefunction interferes at the double slit and is ``guiding'' the particles as
described by Eqn.~\ref{ggbell} and (ii) the initial conditions of the particles
in front of the slits are distributed according to $\rho=|\psi|^2$.  

But what about our earlier claim that Bohmian mechanics allows for the
derivation of Born's rule (i.e. the postulate that $|\psi|^2$ is the
probability density  
for measuring particles in a given volume) rather than postulating it? Antony
Valentini has worked out a 
dynamical explanation \cite{valentini1,valentini2}, how arbitrarily
distributed   
configurations emerge into a $|\psi|^2$ ``equilibrium''. This mechanism has
some resemblance to the corresponding problem in thermodynamics\footnote{There
  exist in fact different strategies to clarify the status of the so-called
  ``quantum equilibrium'', see e.g. \cite{dgz}.}. In any event it is not
possible to prepare a system more precisely than according to the
$\rho=|\psi|^2$ distribution which degrades the determinism of Bohmian
mechanics to an in-principle one\footnote{In this sense the deterministic
  trajectories are like the aether in the Lorentz interpretation
  undetectable}. But that is very much like the situation in 
statistical physics, where 
one still entertains the notion of in-principle deterministic motion although
nobody tries seriously to follow the path of a single particle.  

It should be commented on two peculiarities of the Bohmian trajectories: (i) as
can be seen in Fig.~\ref{ds} they do show completely unclassical behavior like
kinks in the ``field-free'' region between double-slit and screen. But that is
just to say, that they follow a Bohmian and not a Newtonian mechanic. Since
the guiding equation \ref{ggbell} is first order, classical concepts like
momentum, work or energy lose their relevance on the level of individual
trajectories. (ii) Bohmian mechanics is said to be manifestly non-local,
i.e. the motion of any particle is connected to the position of all other
particles, since the wavefunction is defined on the configuration space -- as
opposed to position space. But it is exactly this non-locality, which allows
Bohmian mechanics to violate the Bell inequalities \cite{bell_ungl} 
as demanded by experiment. 
More over the impossibility to prepare systems more precisely than $|\psi|^2$ 
distributed makes sure that signal-locality is obeyed \cite{valentini1}. For
the same reason Bohmian mechanics does also not allow for an experimental
violation of Heisenberg's uncertainty principle \cite{valentini1}. 

For the adherent of Bohmian mechanics it has the virtues to offer a clear 
ontology -- as opposed to the concept of complementarity -- and to provide an
elegant solution to the notorious measurement problem of quantum mechanics
(see e.g. \cite{zurek}). In a nutshell the measurement problem consists of how
to interpret a superposition of macroscopic distinct objects (like pointers,
cats or the like). Ordinary quantum theory has to struggle hard to come up
with a reasonable solution for how to bring together the unitary time
evolution of the Schr\"odinger equation with the seemingly spontaneous
collapse during measurement -- in fact there is no universally accepted
way how to resolve this problem. The poor man's solution being to adopt an
ensemble interpretation of quantum mechanics \cite{statint}, which denies the
applicability of quantum mechanics to individual events. Within Bohmian
mechanics the problem dissolves, since the system is now described by the pair
of wavefunction {\em and} configuration in position space. The continuous
trajectories select the branch of the superposition which will be measured --
an elegant way to get rid of collapse when describing single measurements. 
But again: in the absence of any detailed control over the initial conditions
beyond the $|\psi|^2$ quantum-equilibrium-distribution, this solution of the
measurement problem does not enlarge the predictive power of the theory. 

If one therefore adopts a merely positivistic attitude, Bohmian mechanics and
ordinary quantum theory are completely equivalent since no experiment can
discriminate between them. But people who feel a stronger commitment to
ontology and scientific realism have trouble to come up with a criteria for
which one to prefer\footnote{The serious question of a relativistic
  generalization of Bohmian mechanics is for sure among the important criteria
  for judging the impact Bohmian mechanics. This problem is the subject of
  current research, see  e.g. \cite{bohm_rela}. Similarly the generalization
  of the aether theory to general relativity is problematic.}. 

Bell finishes his paper with a comparison between special relativity and the
Lorentzian aether theory which can be easily adapted to our case:

The approach of Bohr, Heisenberg, Pauli and many others differs from that of
Bohm and de$\,$Broglie in two major ways. There is a difference of
philosophy, and a difference 
in style. The difference in philosophy is this. Since it is experimentally
impossible to predict any single outcome of a quantum process (say, which part
of the screen behind the double slit will be hit next), the standard view
declares probability as a ``irreducible fact of the laws of nature''
(W. Pauli). Bohm, on the other hand, preferred the view that there is indeed a
deterministic substructure, even though the laws of physics conspire to
prevent us from identifying it experimentally. The facts of physics do not
oblige us to accept one philosophy rather than the other. And we need not
accept Bohm's philosophy to accept a Bohmian pedagogy. Its special merit is to
drive home the lesson that it is still {\em possible} to entertain
consistently the notion of quantum
particles moving on deterministic trajectories if we are willing to accept a
non-local dynamic and put away the classical prejudice that these particles
move on a straight path in ``field-free'' space -- free, that is, from fields
other than the de$\,$Broglie-Bohm.  

The difference in style is that instead of inferring quantum phenomena from
known and conjectured laws of physics, the standard presentation has a stronger
emphasis on postulates and axioms. This permits a very 
elegant formulation\footnote{leaving aside for a moment the debatable questions
 whether e.g. the measurement problem can be accounted for convincingly within
 standard quantum mechanics.} as often happens when a few big assumptions can
be made to cover several less big ones. There is no intention here to make any
reservation whatever about the power and precision of the standard approach. 
But in my opinion there is also something to be said for taking students
along the road made by de$\,$Broglie and Bohm\footnote{Among the few modern
  textbooks taking essentially this road is that of James T. Cushing
  \cite{cushing}.}. The longer road sometimes gives more familiarity with
the country.         

In connection with this paper I warmly acknowledge the assistance of 
Roderich Tumulka for teaching Bohmian mechanics to me and Jeremy Butterfield
for valuable comments when preparing this note. 


\end{document}